\PassOptionsToPackage{table}{xcolor}

\documentclass[preprint]{vgtc}               

\usepackage[table,svgnames]{xcolor}
\usepackage{tabularx}

\usepackage{acronym}
\usepackage{balance}

\graphicspath{{figures/}{pictures/}{images/}{./}} 

\usepackage{times}                     

\usepackage{tabu}                      
\usepackage{booktabs}                  
\usepackage{lipsum}                    
\usepackage{mwe}                       

\usepackage{mathptmx}                  

\onlineid{2156}

\vgtccategory{Research}

\title{
    XR Prototyping of Mixed Reality Visualizations:\\
    Compensating Interaction Latency for a Medical Imaging Robot
}

\author{
    Jan Hendrik Plümer
    \thanks{e-mail: jan.pluemer@fh-salzburg.ac.at}\\
    \scriptsize Salzburg University of Applied Sciences\\
    \scriptsize Graz University of Technology
    \and Kevin Yu
    \thanks{e-mail: kevin.yu@tum.de}\\
    \scriptsize medPhoton GmbH
    \and Ulrich Eck
    \thanks{e-mail: ulrich.eck@tum.de}\\
    \scriptsize Technical University of Munich
    \and Denis Kalkofen
    \thanks{e-mail: denis.kalkofen@flinders.edu.au}\\
    \scriptsize Flinders University\\
    \scriptsize Graz University of Technology
    \and Philipp Steininger
    \thanks{e-mail: phil.steininger@medphoton.at}\\
    \scriptsize medPhoton GmbH
    \and Nassir Navab
    \thanks{e-mail: nassir.navab@tum.de}\\
    \scriptsize Technical University of Munich
    \and Markus Tatzgern
    \thanks{e-mail: markus.tatzgern@fh-salzburg.ac.at}\\
    \scriptsize Salzburg University of Applied Sciences
}
\teaser{
  \centering
  \includegraphics[width=.85\linewidth]{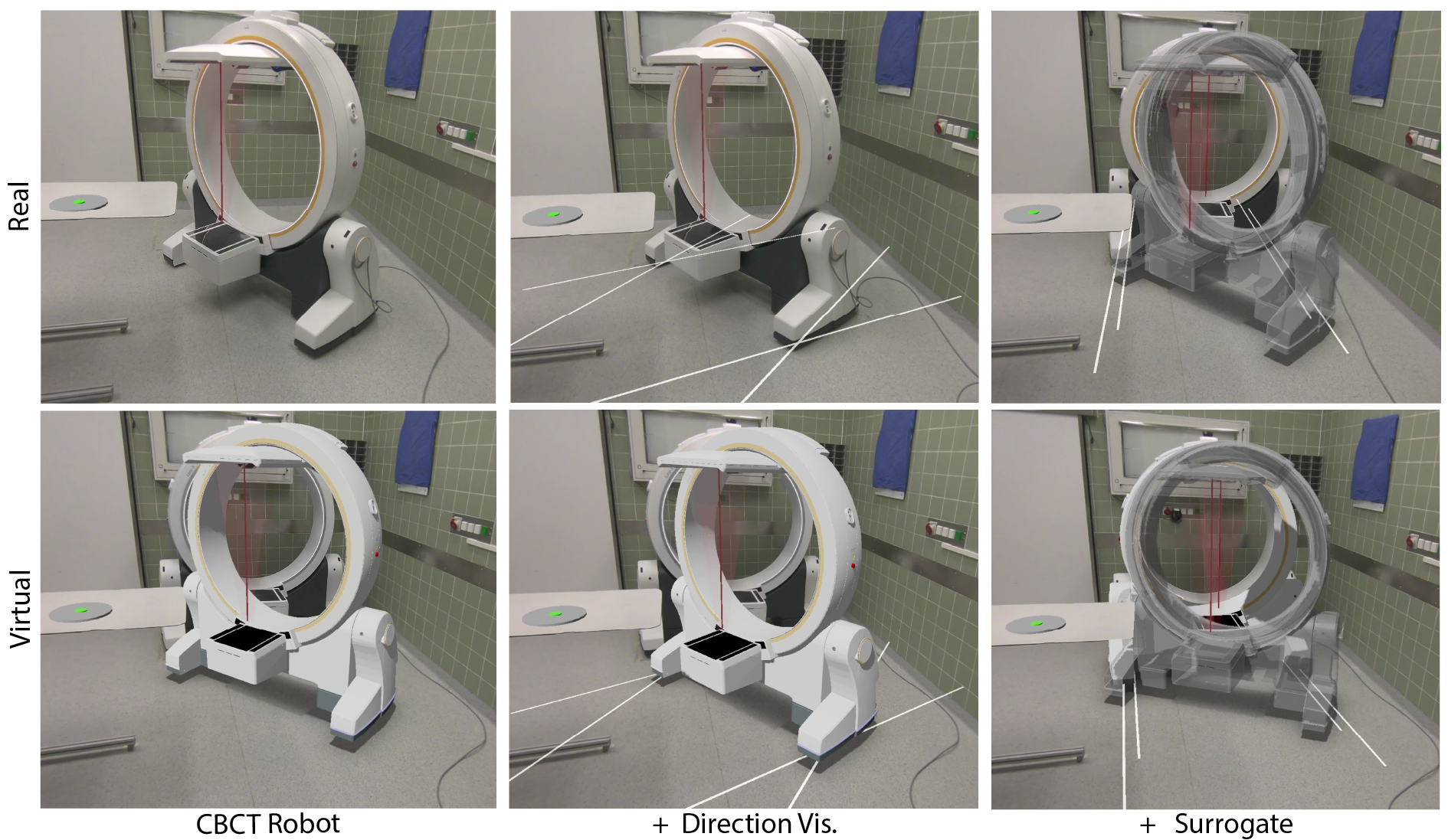}
  \caption{%
  	\textbf{Extended Reality Prototyping (XRP) of Visualizations of Hidden Processes and Surrogates.}     
    We compare interacting with a real mobile imaging robot (Top) against a virtual replica of the same robot (Bottom) in a validation study for Extended Reality (XR) prototyping. 
    When steering the robot, users experience a delay between the user input and the robot actuation due to the robot's internal mechanisms adapting to the user commands. To compensate for this perceived latency, we explore visualizations to improve the responsiveness of the user experience. 
    We compare the robot's off-the-shelf behavior (Left) with two visualizations. 
    (Middle) First, a virtual indicator makes hidden processes of the wheels visible to provide visual feedback and shorten the perceived wait time. 
    (Right) Second, users control a virtual surrogate of the robot instead of the real robot. The real robot follows the surrogate.
  }
  \label{fig:teaser}
}

\abstract{
    Researching novel user experiences in medicine is challenging due to limited access to equipment and strict ethical protocols. 
    \ac{XR} simulation technologies offer a cost- and time-efficient solution for developing interactive systems. 
    Recent work has shown \ac{XRP}'s potential, but its applicability to specific domains like controlling complex machinery needs further exploration.
    This paper explores the benefits and limitations of \ac{XRP} in controlling a mobile medical imaging robot. 
    We compare two \ac{XR} visualization techniques to reduce perceived latency between user input and robot activation. 
    Our \ac{XRP} validation study demonstrates its potential for comparative studies, but identifies a gap in modeling human behavior in the analytic \ac{XRP} validation framework.
}


\newcommand{\condREAL}{REAL}

\newcommand{\condXRP}{XRP}
\newcommand{\condDIR}{DIR}
\newcommand{\condSTE}{STE}
\newcommand{\condSUR}{SUR}

\acrodef{XR}[XR]{Extended Reality}
\acrodef{AR}[AR]{Augmented Reality}
\acrodef{VR}[VR]{Virtual Reality}
\acrodef{MR}[MR]{Mixed Reality}
\acrodef{OST}[OST]{optical see-through}
\acrodef{VST}[VST]{video see-through}
\acrodef{HMD}[HMD]{head-mounted display}
\acrodef{IBR}[IBR]{image-based rendering}
\acrodef{TCT}[TCT]{Task Completion Time}
\acrodef{SEQ}[SEQ]{Single Ease Question}
\acrodef{SUS}[SUS]{System Usability Scale}
\acrodef{XRP}[XRP]{Extended Reality Prototyping}
\acrodef{TCT}[TCT]{task completion time}
\acrodef{CBCT}[CBCT]{cone-beam computer tomography}
\acrodef{SoA}[SoA]{Sense of Agency}

\newcommand{\etal}{et~al.}

\newcommand{\wilc}[3]{Z$=$#1,p#2,r$=$#3}

\newcommand{\art}[4]{$F$(#1,#2)$=$#3,p#4}

\newcommand{\datatable}[3]{#1 & #2 & #3}

\begin{document}


\firstsection{Introduction}
\maketitle

New technologies are continuously integrated into medical applications to improve the effectiveness of procedures and, consequently, patient treatment. 
Recent years have seen a surge of interest in using \ac{MR} for medical applications such as data visualization for surgical planning \cite{tang2018augmented, Song2023}, support during medical procedures~\cite{andress2018fly, qian2017towards,qian2019review}, or 3D medical teleconsultation~\cite{gasques2021artemis, Yu2023, fuchs2014immersive}. 
Research, development, and evaluation of a novel user experience is especially challenging in the medical domain due to the need to follow ethical protocol, e.g., when devices utilize radiation, but also due to a lack of access to the real-world locations, as they typically underlie strict regulations to ensure patient safety. 
Researchers have explored the utilization of \ac{XR} technologies for exploring novel concepts for user experiences under simulated conditions~\cite{Weiss2021, Voit2019, Auer2021, Pluemer2023}, without the need of a corresponding physical realization. 
Such an \ac{XRP} method can lead to more cost-efficient development of interactive systems because qualitative user feedback and quantitative performance can potentially be collected in a simulated environment. 

Previous work on \ac{XRP} has explored the influence of various parameters of the simulation, such as visual realism~\cite{Lee2013}, haptic feedback~\cite{Schott2023a}, and latency~\cite{Lee2010} on user performance. 
Oftentimes, no comparison to a ground truth was provided to establish if results collected in a simulation can be transferred to the real-world use case, and thus the ecological validity of the simulation. 
Other work has designed validation studies~\cite{Makela2020,Mathis2021} that compare a simulated use case against its real-world implementation to move towards establishing the validity of \ac{XRP} methods. 
The results of these provide additional insights to \ac{XRP}, but suffer from confounding factors and make it hard to identify which parameters of the validation study design influenced user feedback and performance. 
To provide structure to \ac{XRP} validation studies, a recently introduced guiding framework~\cite{Pluemer2023} allows researchers and practitioners to identify confounding factors across multiple fidelity dimensions to work towards establishing absolute validity, or relative validity of results gathered in a simulated user experience. 
Absolute validity is established, when users in the simulation and the real use case perform equivalently, and, thus, results can be transferred directly. 
Relative validity means that the relative order effects of, e.g., various interaction methods, behave the same in simulation and real use case. 
The  applicability of the framework was demonstrated in a simple use case controlling a drone in order to explore the impact of functional fidelity, as well as simulation overhead on outcomes of a validation study~\cite{Pluemer2023}. 

In this paper, we explore a more complex user experience involving a user-steerable, mobile medical \ac{CBCT} imaging robot in an \ac{XRP} validation study. 
Based on previous insights from related work on \ac{XRP}~\cite{Pluemer2023, Auer2021, Mathis2021}, instead of showing absolute validity, we aimed to demonstrate relative validity of results by comparing the effect of supporting \ac{MR} visualizations to improve the user experience for a real reference device and its virtual prototype counterpart (see ~\cref{fig:teaser} (Left)). 
In our validation study, we focused on compensating the latency between a user's input and the actuation of the robot to perform an action in order to improve the user experience when controlling heavy machinery or robots. 
Early work for graphical user interfaces showed that such a latency can impact user performance~\cite{MacKenzie1993} which lead to the exploration of latency visualizations such as progress bars~\cite{Myers1985, Harrison2007}, loading screens~\cite{Hohenstein2016} or countdowns~\cite{Ghafurian2020} that had the goal to provide feedback and shorten perceived wait times. 

In a similar vain, we evaluate two visualization techniques for providing visual feedback during human-robot interaction to enable effective interaction. 
The visualizations allow users to understand the current state of the robot and react with necessary adjustments to work towards a goal~\cite{Chen2007}. 
In particular, the first \ac{MR} visualization is visually embedded in the scene and makes hidden mechanical processes of the robot visible with the goal to improve the perceived responsiveness, akin to a progress bar~\cite{Myers1985, Harrison2007}. 
We visualize the mechanics of the robot's rotating wheels when users change direction (see ~\cref{fig:teaser} (Middle)). 
The second visualization allows users to steer a virtual surrogate of the robot, which reacts faster than the original device~\cite{Bejczy1990, walker2019robot} (see~\cref{fig:teaser} (Right)). 
Hence, users can move a robot to a target location, while the real robot follows the path of the surrogate at its slower pace. 
Note that, while we perform a study utilizing a medical imaging robot, the device is representative for other human-controlled machinery and robots suffering from interaction latency, such as industrial arms~\cite{Rakita2020}, or excavators or cranes with extension arms.
In summary, we make the following contributions:
\begin{itemize}
    \item A detailed discussion of a use case for the \ac{XRP} validation framework~\cite{Pluemer2023} demonstrating its benefits when performing and discussing \ac{XRP} validation studies and potential confounding factors. 
    Based on our study, we propose extending the original framework by a discussion of human behavior, as factors such as trust in the robot and the user's sense of agency while steering a self-driving robot, are potential confounding factors that should be considered in \ac{XRP} validation studies.
    \item The validation of \ac{XRP} for the development of \ac{MR} visualizations, demonstrating that relative validity can be achieved with \ac{XRP}. 
    As \ac{XRP} for novel user experiences and ecological validity of results have not been extensively explored, our validation study provides novel insights into \ac{XRP}, thus, pushing the boundaries of  \ac{XRP} for novel product experiences. 
    \item Insights from a comparative study of latency visualization to improve the user experience when interacting with robots and machinery suffering from delays between user input and actuation, making human-robot interaction more effective. 
\end{itemize}

\section{Related Work}
We present related work on frameworks that support the analysis of user experiences, including the \ac{XRP} validation framework, as well as the issue of latency in human-controlled systems.

\subsection{Extended Reality Prototyping}
\ac{XR} simulations are regularly utilized to develop novel products, explore user experiences and implement prototypical applications with the goal to make the design process more time- and cost-efficient and to avoid limitations of current technological state-of-the-art~\cite{Ren2016, Baricevic2012}. 
In product design research, Barbieri \etal~\cite{BARBIERI2013} utilized \ac{MR} to overlay visual designs and interfaces over household appliances to allow users to interact with early designs. 
Other work combined the use of prototypes simulated with \ac{MR} technology with a functional behavior simulation~\cite{bruno2013reliable, bruno2010functional}. 
In HCI, \ac{XR} simulations have been utilized to explore novel authentication procedures~\cite{Mathis2022, Watson2022}, shared interactive spaces~\cite{Jetter2020}, or to explore user behavior in social contexts when interacting with virtual data~\cite{Medeiros2022,Ng2021}. 
Previous work also evaluated the feasibility of novel \ac{MR} visualizations in \ac{XR} simulations, such as advanced adaptive user interfaces~\cite{Cheng2021, Feiyu2022}, guidance visualizations for industrial maintenance~\cite{Burova2020}, or the use of \ac{MR} in urban environments~\cite{Tran2023, Jung2018, Marquardt2020}. 
However, to fully realize the potential of \ac{XR} simulation for product development and prototyping novel user experiences, research must establish ecological validity, i.e., if results gathered in a simulation can be transferred to a real-world implementation of the evaluated scenario. 

Research has followed two approaches to validate the use of \ac{XRP} in order to create simulations that are representative of their real-world counterpart:
Bottom-up, to investigate the effect of distinct system parameters such as latency~\cite{bowman2012evaluating, Lee2009, Lee2010}, or registration error~\cite{bowman2012evaluating, Ragan2009}; or top-down, to implement validation studies that recreate faithful simulations of real-world scenarios in order to compare results of the simulation against its real-world ground truth~\cite{oberhauser2017virtual, Pluemer2023, Voit2019, Weiss2021}. 
However, while qualitative study participant feedback has been similar between both conditions in product design processes~\cite{bruno2010product}, for prototypes of electronic devices~\cite{faust2019mixed, Min2019}, smart artifacts~\cite{Voit2019}, or situated visualizations in \ac{MR}~\cite{Weiss2021}, performance measures typically differed between simulation and real-world use case. 
Reasons were identified in differences between the simulated and the real-world use case, such as technical limitations of the used simulation hardware restricting users~\cite{bruno2010product}, the limited field of view of the used \ac{HMD}~\cite{faust2019mixed}, low resolution \ac{HMD} displays~\cite{Auer2021, Pettersson2019, Savino2019} impacting legibility in the simulation, inaccurate user input~\cite{Mathis2021}, as well as, differences in input modalities~\cite{Makela2020} compared to the real use case, or the behavior of the simulated product not being sufficiently faithful of the original~\cite{Pettersson2019}.

\subsection{Framworks supporting Evaluations}
When designing \ac{XRP} validation studies, it is crucial to align study conditions of the simulation and the real world and control confounding factors, as this increases the chance to collect strong evidence for the utilization of \ac{XRP}, and enables comparability across similar studies.
Previous work proposed various framworks to structure and discuss user studies, e.g., for interaction modalities~\cite{mcmahan2016interaction} or haptic feedback~\cite{Muender2022}. 
For instance, the Human-VE interaction loop~\cite{Bowman2007} distinguishes between different system components and their impact on immersion and therefore on user experience and task performance, and allows to analyse study setups to isolate potential influences on the user experience. 
McMahan \etal~\cite{mcmahan2016interaction} and Münder \etal~\cite{Muender2022} provided detailed frameworks to analyse interaction modalities for \ac{VR}, or haptic feedback modalities. 
Applying the proposed frameworks to design multiple user studies validated them.

Plümer and Tatzgern recently proposed an \ac{XRP} validation framework~\cite{Pluemer2023} based on Human-VE interaction loop~\cite{Bowman2007} to support creating comparable conditions between an \ac{XR} simulation and the real-world use case. 
The framework allows researchers to design validation studies that align study conditions of the simulation and the real world, and to control for confounding factors in these studies. 
The framework structures the interaction with a simulated product by differentiating between input and output devices, as well as the quality of the XR simulation, which includes representations for the user, the environment and the product itself. 
Aside from stimulating the user's senses in terms of visuals, haptics and audio, the framework encompasses interaction fidelity~\cite{mcmahan2016interaction}, and functional fidelity~\cite{alexander2005gaming} of the product, which refers to the behavior of the simulated product in comparison to the real product.
Furthermore, the framework identifies data collection itself as a source of confounding factors, as data collection may be different between real and simulated scenario.

A first application of the \ac{XRP} validation framework~\cite{Pluemer2023} demonstrated that the used simulation hardware, in that case a \ac{VST} \ac{HMD} that was used to visualize the simulated product, introduced simulation overhead that significantly impacted user performance. 
In this paper, we utilize the proposed \ac{XRP} validation framework to set up a comparative study. 
From our study, we derive insights to further refine the validation framework for \ac{XRP}.

\subsection{Latency in Human-controlled Robotic Systems}
Robotic systems operating within complex environments, such as surgical operating rooms, require human intervention for decisional and ethical considerations. 
Interactions between humans and robots involve moving only parts of the robot, such as a single robot arm, or manipulating fully steerable machines. 
Responsiveness and feedback are essential for effective interaction and involves visual and auditory feedback, enabling users to understand the current state of the robot and react with necessary adjustments to work towards a goal. 
Such feedback is especially relevant when the interaction suffers from a temporal delay, or latency, between a user's conceptualization of an action and its execution by the robot. 
Such control latency does not only come from a robot's general inertia due to its construction and involved mechanical parts, i.e., its actuation latency, but also involves sensing and processing latency, which involves communication between sensors, controllers, and actuators of the robot or the processing time for the machine to perceive its environment \cite{falanga2019fast}. 
Remotely controlling robots, i.e. telerobotics, introduce additional temporal delays originating from network communication, which can be measured and optimized to some degree \cite{noguera2023quantifying, Chen2007}. 
Telerobotics provides data on the effect of latency between user input and robot actuation. 
Anvari \etal~\cite{anvari2005impact} reported 500ms latency as the threshold to prevent high error rates. 
Moreover, users perceived a latency of less than 200ms as safe. 
However, a latency of 400-500ms is already tiring for the user, and higher numbers can lead to response times that make robot control infeasible~\cite{xu2014determination}. 
Robots' delayed reactions to user input can also lead to users adapting their interaction strategies which impacts performance and overall user experience as shown in mimicry tasks, where robot arms followed the motion of their users~\cite{Rakita2020, Jensen2015}. 
Multiple works used Machine Learning to predict the movement of the robot to reduce the perceivable latency \cite{andersen2015measuring, sachdeva2021using, farajiparvar2020brief}. 

Previous work also investigated the use of \ac{MR} to reduce latency through virtual surrogates, where users steer a virtual copy of a robot that does not suffer from any input latency~\cite{Chen2007}. 
An early example of such a technique is the ``phantom robot''~\cite{Bejczy1990} where operators control a virtual copy overlaid over a video feed of the real robot in a simple tapping task to compensate for the delay between user input and execution of an action by the robot. 
Preliminary evaluation showed promising results to improve operator performance for delayed inputs. 
Hashimoto \etal~\cite{hashimoto2011touchme} also utilize such a transparent virtual surrogate of a robot and overlay it on the camera stream looking at their robot system. 
Users could manipulate the surrogate to plan the motion of the robot. 
They found that sequential movement execution after motion planning using the surrogate is more efficient than simultaneous actuation and user input. 
Walker \etal~\cite{walker2019robot} showed that abstracting the control of a robotic system with a virtual surrogate can make motion planning, in their case, a drone, more intuitive for the user. 
Their system then automatically adjusted parameters for path planning and fine controlling. 
Similar to them, Luz \etal~\cite{luz2023enhanced} utilized a predictive visualization with direct input with a virtual surrogate to compensate a three-second network latency. 
In their evaluation, they found that the virtual surrogate yielded higher usability and a decreased workload for the user. 
Users preferred the virtual surrogate, especially for long-distance motion planning, and the predictive visualization for short precision movements. 
We also utilize a virtual surrogate to compensate for a mobile imaging robot moving at relatively slow pace, with the goal to make interaction more efficient. 
In contrast to previous work, we compare two types of latency visualizations: a latency visualization that visualizes internal, invisible processes of a robot, akin to a progress bar~\cite{Myers1985, Harrison2007} that shows users that the input has an immediate effect, and a surrogate visualization, where users steer a more responsive copy of the robot and, thus, receive immediate feedback to their input~\cite{Bejczy1990, walker2019robot}. 
Furthermore, we utilize these visualizations to explore the prototyping of novel user experiences with \ac{XRP} as we aim to to demonstrate relative validity between conditions in an \ac{XRP} validation study.

\section{Mobile Imaging Robot}
We utilize a mobile \ac{CBCT} robot, the Loop-X (medPhoton GmbH, Austria), as the product for an \ac{XRP} validation study in order to explore the benefits of utilizing \ac{XRP} for product development. 
This robot is representative for complex, highly innovative products for which it is costly and time-consuming to prototype and iterate over variations of the design and user experience. 
Hence, our evaluation can be performed with any similar fully robotic systems and is not limited to our robot. 
Evaluating the user experience is particularly difficult for medical devices as in-depth evaluations underlie strict regulations and access to the target environment is often restricted. 
The \ac{CBCT} robot also is representative for other heavy machinery that typically suffers a delay between user input and robot actuation performing the desired actions, such as truck-mounted cranes, excavators with extendable arms, or industrial robot arms. 
In the following, we provide details of the used mobile \ac{CBCT} scanner and its digital version that is used in the presented \ac{XRP} validation study.

\subsection{Real Device}
The mobile robot is tailored for X-ray and \ac{CBCT} image-guided medical procedures (e.g. spine surgery, brachytherapy) and supports automated and human-guided translational and rotational movements across the floor and independent adjustments of the gantry tilt, X-ray source and detector arm. 
The robot weighs approximately 500 kilograms. 
The velocity of the robotic system's components is limited by actuation time, frequently constrained by the specifications of its electric motors. 
Additionally, as human operators are nearby, additional safety criteria come into play that limit the speed of movement based on the situation. 
Consequently, the maximum default velocity is limited to a translation speed of 10cm/s in clinical mode.
Furthermore, four wheels on the bottom side of the robot control the direction of movement, and the system needs to align them to the desired direction, enabling the robot to rotate around its center with a rotation speed of 7.5°/s.
Due to physical constraints, the rotational speed of the wheels for aligning them with the driving directions is capped at 22°/s. 
Therefore, the prevalent latency between user input, start of the movement, and arrival at the desired goal position is the actuation time of the driving motors and wheel rotations. 
Hence, input and action may have a perceived actuation latency of several seconds, depending on the directional change.
A portable control panel of about 2.55kg based on a commercially available tablet device and additional joystick inputs communicate the control inputs of the machine through a wireless network generated by the Loop-X via safe TCP channels. 
Given the co-located spatial proximity of the control panel and the robot, network latency remains below three milliseconds, rendering it negligible for our investigations. 
Moreover, unlike human-like teleoperated robots, creating agency for the user is unnecessary as the robot does not resemble humanoid motions. 
We emphasize that the latency compensation methods investigated in this paper could be applied to other machinery suffering from input and actuation latency.

\subsection{Digital Version}
To create a virtually identical  digital version of the Loop-X in terms of visual appearance, we utilized existing CAD data created during the planning stage of the device. 
To achieve a high level of realism for the virtual robot the polygon count was reduced to 566k triangles preserving details during rendering.
To achieve the same behavior for the digital version of the mobile robot, when users steer the ring, the simulated system receives the same actuation commands than the real ring to adjust the rotation of the digital robot's virtual wheels. 
Users controlled the digital version of the ring using the same physical control panel. 
The user input was sent to the Unity engine via wireless TCP connection, and translated to wheel manipulations and translational, as well as rotational movement of the robot with a wheel alignment rotation speed of 22°/s, robot translation speed of 10cm/s, and speed to rotate the robot around its center of 7.5°/s.

\begin{figure*}[t!]
    \centering
    \includegraphics[width=.9\textwidth]{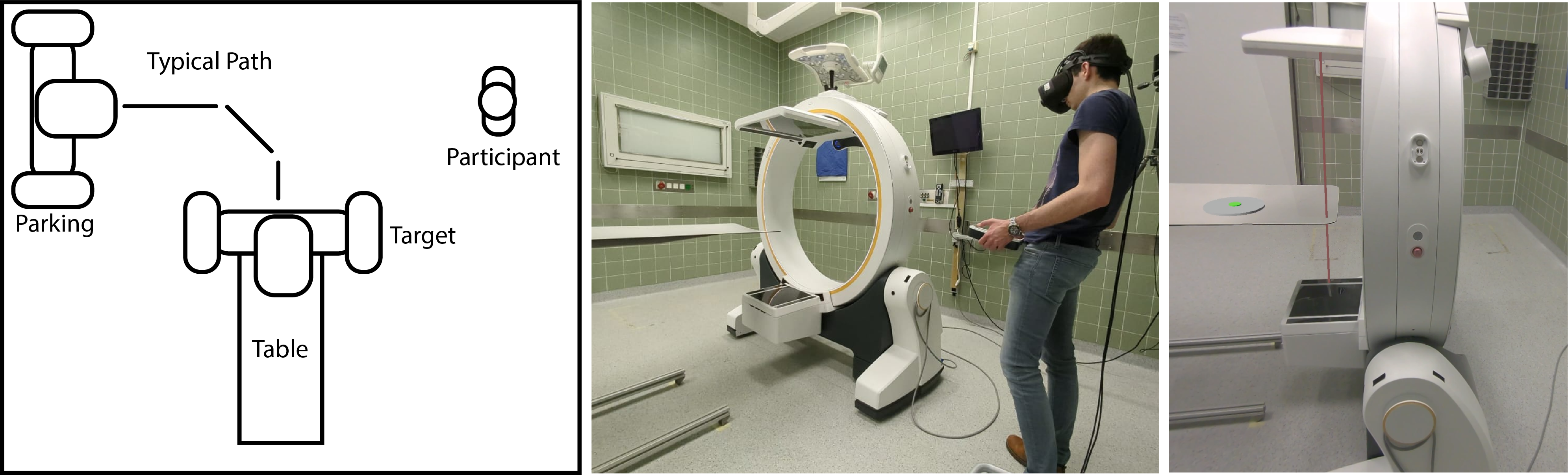}
    \caption{
        \textbf{Apparatus.} (Left) A sketch of the room setup with ring and surgical table. The line indicates a typical path of users steering the mobile robot to the target location at the table. (Middle) The room with a participant steering the mobile robot and holding the controller. (Right) A view of the \ac{MR} laser augmentation that users had to align with the target location at the surgical table indicated by the augmented grey and green circle. 
    }
    \label{fig:apparatus}
\end{figure*}

\subsection{Mixed Reality Visualizations}
For our \ac{XRP} validation study, we explored supporting \ac{MR} visualizations to improve the handling of the mobile robot. 
A basic visualization made an integrated laser visible in order to support the alignment of the robot with a target location (see \cref{fig:apparatus} (Right)). 
However, our main focus was to explore the effect of latency visualizations in terms of objective and subjective user performance. 
A main issue for the delayed actuation of the robot is the alignment of the mechanical wheels which dictate the driving direction, an internal operation that is typically hidden from the user. 
Therefore, we introduced a visualization that showed the current alignment of the wheels as lines indicating the driving direction.
When the user changed direction, the wheels and, thus, the lines would realign and provide instant feedback to users as well as visualize the progress of the performed operation (see \cref{fig:teaser} (Middle)). 
Graphical user interfaces utilize loading screens~\cite{Hohenstein2016}, progress bars~\cite{Myers1985, Harrison2007} or countdowns~\cite{Ghafurian2020} for a similar purpose. 
As the line visualization still required users to wait until wheels would realign before the ring started moving, we also created a visualization that combined the line visualization with a surrogate visualization~\cite{walker2019robot, Bejczy1990}, where users would steer a copy of the robot that reacted instantaneously to user input (see \cref{fig:teaser} (Right)). 
The actual robot moving at their regular pace followed this surrogate ring until the target location.

\section{XR Prototyping of MR Visualization}
We follow the proposed structure of the \ac{XRP} validation framework~\cite{Pluemer2023} to describe the study and discuss potential confounding factors, i.e., differences between real and simulated use case. 

\subsection{Study Design}
We designed a within-subject user study to explore whether latency visualizations can facilitate the steering of a slow-moving robot. We introduced two independent variables: Realism and Visualization. 

\textbf{Realism} had two conditions: a reference condition where participants interact with the real-world robot (\textbf{\condREAL{}}) and an \ac{XR} condition (\textbf{\condXRP{}}) where users controlled a simulated robot version. 

\textbf{Visualization} had three conditions: 
Direct control (\textbf{\condDIR{}}), where the participants steered the robot directly. In this condition, the robot reacted to user input as it realigned its hidden mechanics (i.e., wheels) to follow user commands and moved into the intended direction. 
Secondly, direct control with a supporting steering visualization (\textbf{\condSTE{}}), where the steering is the same as in \condDIR, but the typically hidden motion of the robot's mechanics reacting to user input was visualized. 
Hence, each of the four robot wheels were visually augmented by a line on the ground indicating the current alignment of the respective wheel. Wheel and, thus, the line immediately reacted to user input even though the robot itself did not start its movement. 
The third condition was surrogate control (\textbf{\condSUR{}}), where the user steered a copy of the robot that did not suffer from latency. Depending on the given Realism condition, either the real robot, or the simulated robot followed the exact same path autonomously. 
To distinguish between the robot and its surrogate representation, the surrogate robot was partially transparent.


\textbf{Task.} 
While the robot is capable of acquiring cone beam CT images, we only utilized the robot's basic steering functionality for a task.
Hence, participants perform a meaningful task close to a real use case in a realistic setting, in our case a medical robot in a surgery theater, while at the same time providing a task that is representative of other self-driving and human-steered robots, in order to ensure generalizability to other shapes of robots. 
Participants steered the robot to a target location and aligned it with a target. 
Participants always started at the parking position and moved the robot towards a target location on a surgical table. As we restricted the movement area of the robot, participants had to utilize a combination of translational and rotational commands to move the robot to the target location (see \cref{fig:apparatus} (Left and Middle)). Simultaneous movements with rotational steering were not available for the users.  To align the robot with the target, participants relied on a visualization that made the targeting laser of the robot visible. 
Alignment was finished, when the laser was positioned inside the target area (see \cref{fig:apparatus} (Right)). Both the target and the beam were virtual. In \condSUR{}, participants aligned the surrogate, while the robot was slowly approaching its target location.
Upon reaching the target location, participants were asked to move the robot back to its parking position to extend the exposure to each condition.

\definecolor{FidelityBlue}{RGB}{145,191,219}
\definecolor{FidelityYellow}{RGB}{255,255,191}
\definecolor{FidelityRed}{RGB}{252,141,89}

\newcolumntype{Y}{>{\centering\arraybackslash}X}
 
\begin{table*}[t!]
    \begin{tabularx}{\textwidth}{|l|r|X|X|}
      \hline
      \multicolumn{2}{|c|}{Fidelity} & \condREAL{} & \condXRP{} \\
      \hline
            Visual
            & Product 
            & \cellcolor{FidelityBlue} real robot
            & \cellcolor{FidelityYellow} simulated robot
            \\ 
            
           \cline{3-4}
           & Control 
           & \multicolumn{2}{>{\hsize=\dimexpr2\hsize+2\tabcolsep+\arrayrulewidth\relax}c}{\raggedright\cellcolor{FidelityBlue} real control device}
           \\
           
           \cline{3-4}
           & Environment 
           & \multicolumn{2}{>{\hsize=\dimexpr2\hsize+2\tabcolsep+\arrayrulewidth\relax}c}{\raggedright\cellcolor{FidelityBlue} real environment with registered phantom geometry for 
           occlusion of augmented information (e.g. surrogate)}
           \\
           
           \cline{3-4}
           & User
           & \multicolumn{2}{>{\hsize=\dimexpr2\hsize+2\tabcolsep+\arrayrulewidth\relax}c}{\raggedright\cellcolor{FidelityBlue} real user}
           \\   
           
           \cline{3-4}
           \hline\hline
            \multicolumn{2}{|l|}{Haptic}
            & \multicolumn{2}{>{\hsize=\dimexpr2\hsize+2\tabcolsep+\arrayrulewidth\relax}c}{\raggedright\cellcolor{FidelityBlue} same haptic feedback from the handheld control panel }
            \\
           
           \cline{3-4}
           \hline\hline
            \multicolumn{2}{|l|}{Audio}
            & \multicolumn{2}{>{\hsize=\dimexpr2\hsize+2\tabcolsep+\arrayrulewidth\relax}c}{\raggedright\cellcolor{FidelityBlue} real audio of handheld control panel}
            \\
           
           \cline{3-4}
           \hline\hline
            \multicolumn{2}{|l|}{Interaction}
            & \multicolumn{2}{>{\hsize=\dimexpr2\hsize+2\tabcolsep+\arrayrulewidth\relax}c}{\raggedright\cellcolor{FidelityBlue} real control device }
            \\
           
           \cline{3-4}
           \hline\hline
            \multicolumn{2}{|l|}{Functional}
            & \cellcolor{FidelityBlue} real tracked robot
            & \cellcolor{FidelityYellow} robot simulation based on real robot's control mechanisms, leading to same behavior and speed. deviations in behavior from physical motion differences (e.g., slipping wheels)
            \\
           
           \cline{3-4}
           \hline\hline
            Data
            & Objective
            & \multicolumn{2}{>{\hsize=\dimexpr2\hsize+2\tabcolsep+\arrayrulewidth\relax}c}{\raggedright\cellcolor{FidelityBlue} same automated TCT collection for all conditions}
            \\

            \cline{3-4}
           & Subjective 
           & \multicolumn{2}{>{\hsize=\dimexpr2\hsize+2\tabcolsep+\arrayrulewidth\relax}c}{\raggedright\cellcolor{FidelityBlue} same questionnaires for all conditions}
           \\
           
           \cline{3-4}
           \hline\hline
            \multicolumn{2}{|l|}{Simulation Overhead}
            & \multicolumn{2}{>{\hsize=\dimexpr2\hsize+2\tabcolsep+\arrayrulewidth\relax}c}{\raggedright\cellcolor{FidelityBlue} same \ac{HMD} for all conditions}
            \\

      \hline
    \end{tabularx}
    \caption{\textbf{Fidelity Analysis of the Two Study Conditions}. Alignment between the simulated condition (\condXRP{}) and the real-world scenario (\condREAL{}) is indicated in \colorbox{FidelityBlue}{blue}. Fidelity differences unique to the \ac{XR} simulation are indicated in \colorbox{FidelityYellow}{yellow}.}
    \label{tab:fidelity_analysis}
\end{table*}

\textbf{Data Collection.} 
We aimed to support generalization of the validation study results by including dependent variables that are typically utilized in performance evaluations of products~\cite{Pluemer2023}. 
For subjective data, we measured the task load with the raw NASA TLX~\cite{HART1988} and usability with the \ac{SUS}~\cite{Brooke96sus} after each Visualization condition.  
After finishing a Realism condition, we asked the participants to rank the visualization conditions for preference. 
We also performed a semi-structured interview after finishing all conditions. 
Concerning objective data, we measured \ac{TCT} for reaching the target location, beginning when participants activated the ring using a dedicated button on the control panel and ending when the participant confirmed to have reached the target position.
All measurements were automatically calculated within the simulation software. 
Instead of measuring \ac{TCT} in \condSUR, we measured TCT$_{SUR}$ as the time required to position the surrogate visualization as the task was completed by participants and the ring had to autonomously drive to the target location. 


\textbf{Apparatus.} For all conditions, we utilized the \ac{VST} mode of a Varjo XR-3 \ac{HMD}. 
The simulation was implemented in the Unity engine and ran on a PC with an NVIDIA GeForce RTX 3080 Ti graphics card, AMD Ryzen 7 5700X CPU and 32 GB of RAM. 
We utilized the Loop-X mobile \ac{CBCT} robot for the \condREAL{} condition, and a virtual copy of it for the \condXRP{} condition. In both conditions, participants used the original handheld remote control panel of the robot.
Positional tracking of the real device was achieved with one Vive Tracker (v2018) mounted on top of it and Vive Base Stations (v2.0).
To create a virtual representation of the study room to provide occlusion geometry for virtual geometry, the real room was captured using the 3D scanner App\footnote{\url{https://apps.apple.com/us/app/3d-scanner-app/id1419913995}, last accessed March 11th, 2024}, installed on an iPhone 13 Pro Max.

\subsection{Simulation Fidelity}
We analyse the fidelity of the \ac{XRP} validation study (see \cref{tab:fidelity_analysis}) based on the framework proposed in~\cite{Pluemer2023} to identify similarities and confounding factors between the Realism conditions. 

\textbf{Visual Fidelity.}
Participants wore the \ac{HMD} during all conditions utilizing its \ac{VST} capabilities, which lead to users perceiving themselves, the environment and the control panel mediated through the \ac{VST} camera. Hence, no virtual visual representation was required for user, environment and control panel.
Key features of the environment, such as the surgical table, had an invisible virtual representation. They occluded virtual elements such as the surrogate and robot's virtual representation, aiding users in understanding the spatial relations of the scene correctly.
In \condREAL{}, participants perceived the real robot mediated through the cameras of the \ac{HMD}. 
To digitally recreate the appearance of the robot, we utilized precise CAD model of the real robot. 
Laser beam, target marker and the virtual surrogate were augmentations and, thus, purely virtual in both Realism conditions. 
The light setup was controlled. We prevented direct and indirect sunlight by darkening the room and relying solely on the artificial ceiling lights. Lights lead to shadows directly below the real device, which we simulated for the virtual devices as well. 
Overall, preliminary feedback found that the visual representation of the robot was very similar between both Realism conditions. 

\textbf{Haptic Fidelity.}
Haptic fidelity was the same in \condREAL{} and \condXRP{}, as participants utilized the same control panel in both conditions. 
Participants also did not interact with virtual geometry or the robots themselves, thus, no additional haptic feedback was required.

\textbf{Audio Fidelity.}
During the actuation of the mobile robot, including the duration of the realignment of the hidden wheels, the handheld control panel provided auditive user feedback in the form of beeping sounds to indicate that the robot was moving. 
This auditive feedback played during all experimental conditions, including conditions of \condXRP{} where users only controlled the virtual robot. When users changed the movement direction of the actual robot, they could faintly hear the motors for realigning the steering wheels.
Further, the robot additionally produced a constant humming background noise due to its active ventilation. 

\textbf{Interaction Fidelity.} Interaction fidelity was the same in \condREAL{} and \condXRP{}, as both conditions utilized the same control panel.

\textbf{Functional Fidelity.} 
We validated the approximation of the simulation by comparing the speeds of the digital ring used in \condXRP{} against the real robot of \condREAL. 
Both the real and virtual robots moved at a translation speed of 10cm/s, rotated at an angular speed of 7.5°/s, and wheels aligned with the driving direction at 22°/s. 
Further, the real robot has a short acceleration and deceleration phase of approximately 4 centimeters due to inertia, which we modeled accordingly for the virtual robot. 
Similar to visual fidelity, preliminary feedback suggested that both versions behaved similarly. 
However, previous work determined that achieving full functional fidelity, i.e., the same behavior is unlikely~\cite{Pluemer2023}. 
For instance, the real robot occasionally suffers from slipping wheels on the floor when it loses traction. 
Such unpredictable physical behavior could not be replicated in our virtual model and, and thus, absolute validity where virtual and real devices behave the same is outside the scope of this study. 
Therefore, we focused on exploring \ac{XRP} to achieve relative validity by comparing prototyping variations of a product.

\textbf{Data Fidelity.}
Subjective data collection was the same in both \condREAL{} and \condXRP{} as we utilized the same questionnaires. 
Furthermore, we maintained identical approaches for objective data collection (\ac{TCT}) for all data measurements through automated software routines regardless of the condition. 

\textbf{Simulation Overhead.}
We designed the study in a way that wearing an \ac{HMD} was part of all conditions in both \condREAL{} and \condXRP{} in order to effectively eliminate simulation overhead of the \ac{HMD} as a confounding factor~\cite{Pluemer2023}. 
We made the \ac{VST} \ac{HMD} an integral part of the product experience and provided latency and surrogate visualizations, as well as basic visualizations of the laser and target location to improve the general user experience.


\section{Experiment}
We describe the experiment based on the presented study design.

\textbf{Participants.} 24 participants were recruited from campus announcement and local public announcements (age=25.5 (sd=3.66) min=18 max=36, female=12, male=12) and received a small monetary compensation for their efforts. 
All participants had normal or corrected to normal vision and used digital media on daily basis with an average of 48.8 (sd=26.25, median = 50) hours per week, of which an average 3.2 (sd=4.57, median = 0) hours are video games.
22 participants had a computer science or robotics background, of which 5 additionally had a medical background. One participant was a medical student. 
Four participants had no VR/MR experience, only two participants used these technologies on a weekly basis. 
Four participants were aware of our robot in our study beforehand and interacted with it at least once before the study. Only one of the participants stated their accumulated usage time with the robot exceeded ten hours.

\textbf{Procedure.} The institutional ethics committee of the Salzburg University of Applied Sciences approved this study. 
At the time of the experiment, participants signed an informed consent and filled out a demographic questionnaire. 
Participants were instructed that they could remove the \ac{HMD} or quit the experiment at any time. 
The condition Realism was counterbalanced, so that half of the participants started with \condREAL{}, the other half with \condXRP{}. 
The condition Visualization was balanced using a 3x3 Latin Square Table. 
The study started once participants had been familiarized with the \ac{HMD} and the control panel and the steering of the robot. 
After finishing a task in a Visualization condition, participants filled out questionnaires on NASA-TLX and SUS, resulting in 16 Likert-scale questions.
After finishing all Visualization conditions for a Realism condition, participants ranked the Visualizations and afterwards continued with the next Realism condition. 
This process was repeated for all conditions, the experiment ended with a semi-structured interview. 
We collected 2 (realism) x 3 (visualization) = 6 data points per participants and, overall, 144 data points over all participants.
Participants required on average 75 minutes to complete the study, with the longest duration observed at 90 minutes including introduction and interview phases.

\textbf{Hypotheses. } 
\textbf{H1.} We expected \condDIR{} to perform worse than \condSTE{}, and \condSTE{} to perform worse than \condSUR{} in terms of performance and preference (i.e., \condDIR{} < \condSTE{} < \condSUR{}), as the visualizations in \condSTE{} and \condSUR{} add a level of responsiveness to an otherwise less agile system with inherent latency.

\textbf{H2.} Despite our best efforts to align the behavior of the controlled devices in \condREAL{} and \condXRP{}, we expected that the results would show performance differences between the same visualization conditions in \condREAL{} and \condXRP{}, and, thus, not lead to absolute validity~\cite{Auer2021, bruno2010product, faust2019mixed}. While the simulation allowed us to create a digital copy that steers the same as the real robot, we expected that differences in functional fidelity would still lead to performance differences~\cite{Pluemer2023, Mathis2021}. 

\textbf{H3.} While the same Visualization conditions in \condREAL{} and \condXRP{} will show performance differences (see H2), the relative effects and, thus, performance differences of the Visualization conditions in a Realism condition will be the same for both \condREAL{} and \condXRP{} and follow the order as outlined in H1 (\condDIR{}  < \condSTE{} < \condSUR{}).

\subsection{Results}
\begin{figure}[tbp]
    \centering
    \includegraphics[width=.95\columnwidth]{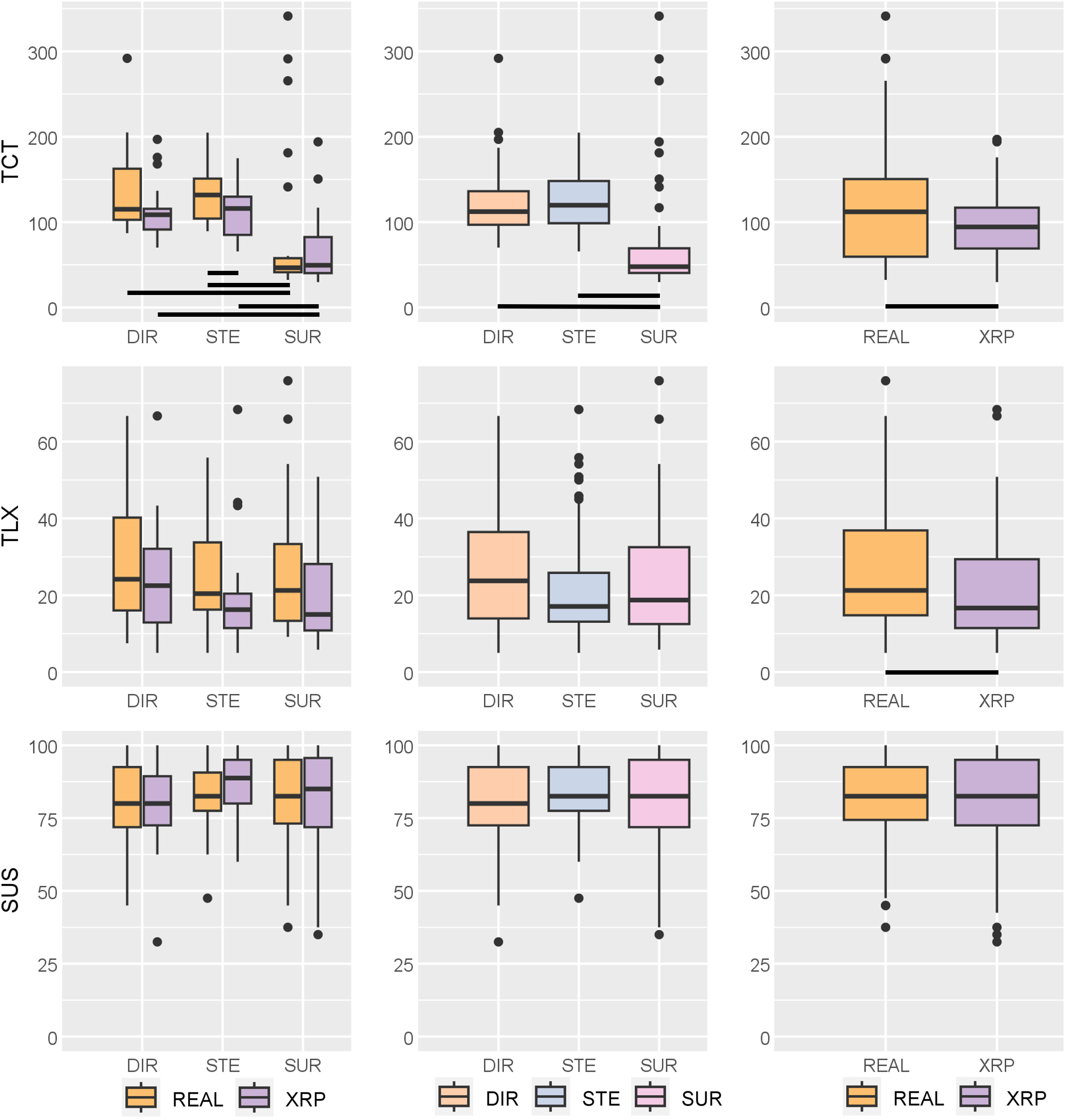}
    \caption{
        \textbf{Measures} of Task Completion Time (TCT, seconds), TLX and SUS for Visualization and Realism conditions. Lines indicate statistically significant differences.
    }
    \label{fig:overalltctsustlxplot}
\end{figure}

\begin{figure}[h]
    \centering
    \includegraphics[width=.95\columnwidth]{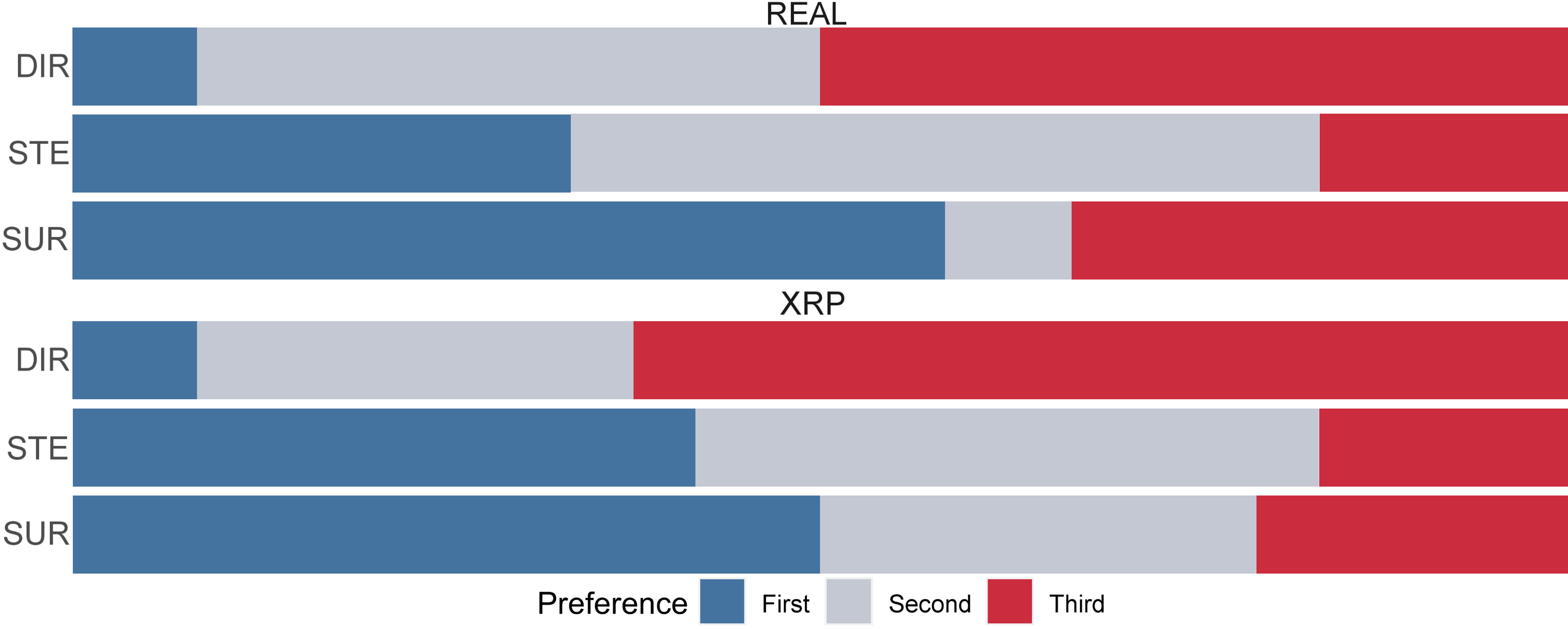}
    \caption{
        \textbf{Preference Ranking} for both Realism conditions.
    }
    \label{fig:preference}
\end{figure}

The statistics software \emph{R} was used, data was evaluated with a significance level of 0.05. The data residuals did not fulfill the normality requirement. Therefore, we utilized align-and-rank transform (ART)~\cite{Wobbrock2011} tests and follow-up Wilcoxon signed-rank tests for post-hoc analysis. The reported p-values are Bonferroni-Holm corrected. We calculate effect size for Wilcoxon signed-rank tests as $r=\frac{Z}{\sqrt{N}}$~\cite{Fritz2012}. 
Results are visualized in~\cref{fig:overalltctsustlxplot},~\cref{fig:preference} and~\cref{fig:boxplots_overall}. 
More details such as complete descriptive statistics can be found in the supplemental material. The  \ac{TCT} measurements averaged over all participants for the respective conditions shown in the box plots are presented in ~\cref{tab:descriptivestatsconditions}.



ART revealed statistically significant difference for \textbf{Mental Demand} in realism (\art{1}{23}{10.87}{=0.003}), for \textbf{Physical Demand} in visualization (\art{2}{46}{4.04}{=0.024}) and in realism:visualization (\art{2}{46}{3.47}{=0.040}), for \textbf{Success} in realism:visualization (\art{2}{46}{5.83}{=0.006}), for \textbf{Commitment} in realism (\art{1}{23}{5.04}{=0.035}), for \textbf{Stress} in realism (\art{1}{23}{7.19}{=0.013}) and in visualization (\art{2}{46}{8.28}{<.001}), for \textbf{TCT} in realism (\art{1}{23}{5.54}{=0.027}), in visualization (\art{2}{46}{39.80}{<.001}) and in realism:visualization (\art{2}{46}{7.82}{=0.001}), and for \textbf{Task Load} in realism (\art{1}{23}{7.62}{=0.011}).


\begin{figure}[h]
    \centering
    \includegraphics[width=.95\columnwidth]{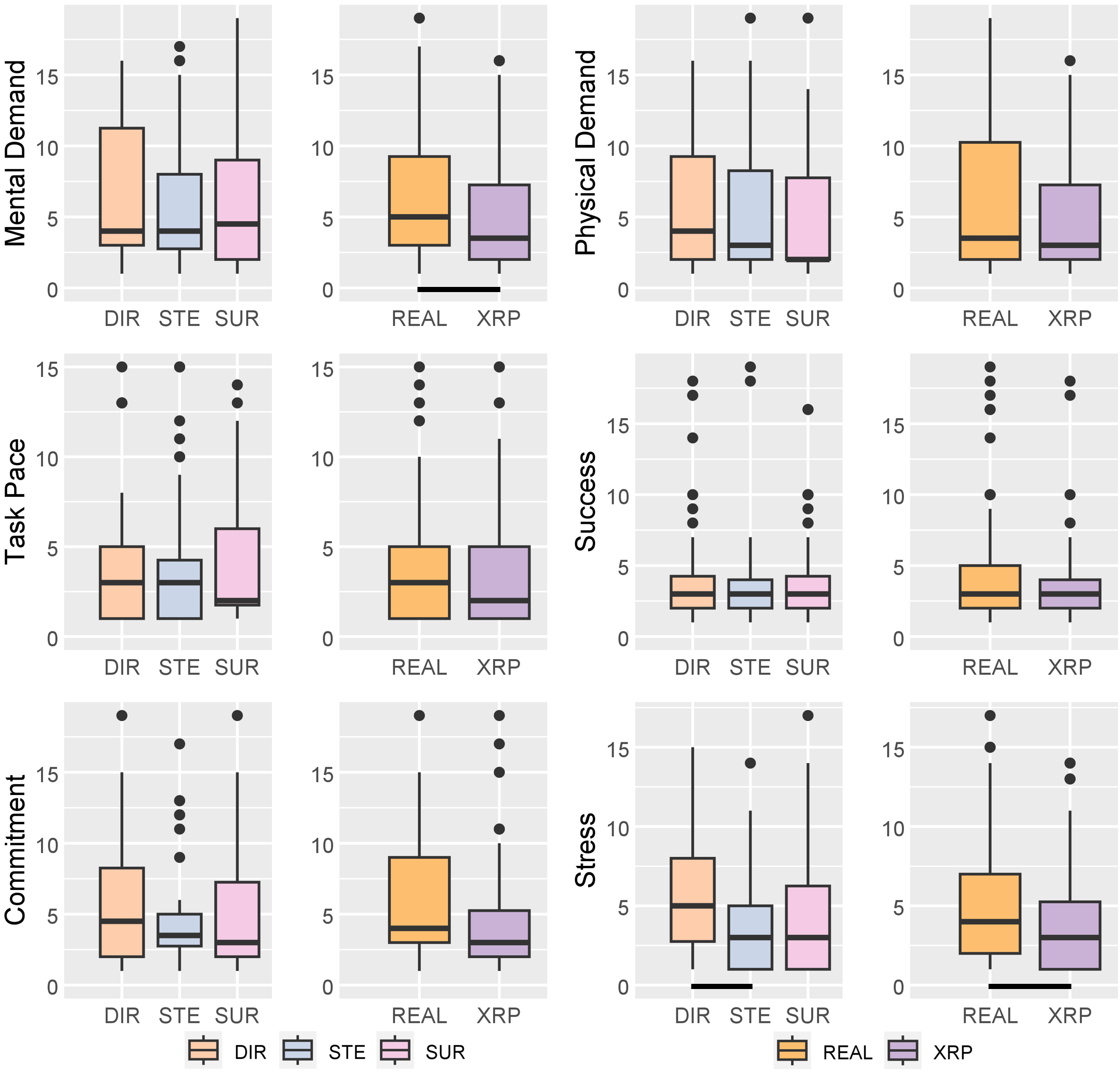}
    \caption{
        \textbf{Box Plots for Overall Visualizaton and Realism Conditions (NASA TLX)}. Lines indicate statistically significant differences. 
    }
    \label{fig:boxplots_overall}
\end{figure}

Follow-up pairwise tests after calculating Bonferroni-Holm corrected p-values revealed no sign. difference for \textbf{Physical Demand}, \textbf{Success} and \textbf{Commitment}, significant differences for \textbf{Mental Demand} between \condREAL{} and \condXRP{} (\wilc{3.19}{=0.001}{0.46}), significant differences for \textbf{Stress} between \condDIR{} and \condSTE{} (\wilc{3.28}{=0.003}{0.47}) and between \condREAL{} and \condXRP{} (\wilc{2.20}{=0.028}{0.32}), significant differences for \textbf{TCT} in \condREAL{} between \condDIR{} and \condSUR{}  (\wilc{3.09}{=0.014}{0.45}), and between \condSTE{} and \condSUR{} (\wilc{2.66}{=0.047}{0.38}), in \condXRP{} between \condDIR{} and \condSUR{} (\wilc{3.66}{=0.002}{0.53}), and \condSTE{} and \condSUR{} (\wilc{3.60}{=0.003}{0.52}), between \condREAL{} and \condXRP{} for \condSTE{} (\wilc{2.60}{=0.047}{0.38}), between \condDIR{} and \condSUR{} (\wilc{3.11}{=0.006}{0.45}), between \condSTE{} and \condSUR{} (\wilc{3.00}{=0.005}{0.43}) and between \condREAL{} and \condXRP{} (\wilc{2.17}{=0.03}{0.31}), and significant differences for \textbf{Task Load} between \condREAL{} and \condXRP{} (\wilc{2.96}{=0.003}{0.43}).

\begin{table}[h]
    \centering
    \begin{tabular}{|c|c|ccc|}
      \hline
      Real. & Vis. & Mean & Std.Dev & Median \\
      \hline
       & \condDIR   &  \datatable{135.25}{47.72}{115.0} \\
      \condREAL & \condSTE   & \datatable{133.06}{31.08}{131.8} \\
       & \condSUR  & \datatable{86.25}{89.72}{46.6} \\
      \hline
       & \condDIR &\datatable{111.33}{31.83}{108.6}\\
        \condXRP & \condSTE &\datatable{112.54}{30.18}{115.9}\\
         & \condSUR & \datatable{66.85}{40.04}{49.5}\\
        \hline
         \condREAL &  &\datatable{118.19}{64.62}{112.0}\\
         \condXRP & &\datatable{96.91}{40.01}{94.3}\\
         \hline
         &\condDIR   & \datatable{123.29}{41.91}{112.3}\\
         &\condSTE   & \datatable{122.80}{32.03}{119.8}\\
         &\condSUR   & \datatable{76.55}{69.42}{47.8}\\
         \hline
    \end{tabular}
    \caption{\textbf{Descriptive Statistics} of Task Completion Time (seconds).}
    \label{tab:descriptivestatsconditions}
\end{table}

\section{Discussion}
In the following, we discuss the hypotheses and derive insights for further \ac{XRP} validation studies and latency visualization methods.

\subsection{Hypotheses}

\textbf{H1.} 
As expected, we found significant differences between Visualization conditions. Overall, \ac{TCT} (see ~\cref{fig:overalltctsustlxplot} (Top)) results show that \condSUR{} was significantly faster than both \condDIR{} and \condSTE{}. 
In terms of task load, latency visualization also significantly reduced stress for participants between \condDIR{} and \condSTE, and \condSUR{} was also lower than \condDIR{}, though not statistically significant.
Overall, task load was lower for conditions with latency visualization (see ~\cref{fig:overalltctsustlxplot} (Middle)), although not statistically significant. 
Latency visualizations were preferred by participants, with overall 54\% ranking \condSUR{} in first place and 38\% ranking \condSTE{} in second place (see ~\cref{fig:preference}). In only 8\% of all cases, participants ranked the direct control (\condDIR{}) in first place. In the post-experiment interview, 79\% of participants explicitly mentioned that the surrogate visualization (\condSUR) was helpful, and 58\% mentioned the same for the direction visualization (\condSTE). 
Overall, we partially accept H1 as the \condSUR{} latency visualization has a beneficial effect on the users' performance. 
 
\textbf{H2.} %
As expected, based on previous experiments in literature~\cite{Pluemer2023,Auer2021,Mathis2021}, we detected differences between the Realism conditions with \condXRP{} being significantly faster than \condREAL{}. 
Hence, despite our best efforts, the conditions were apparently not completely aligned. 
Our fidelity analysis (see ~\cref{tab:fidelity_analysis}) indicates that the main objective difference between conditions is the functional fidelity of the ring, i.e., its behavior is not exactly represented in the \condXRP{} condition, even though 67\% of participants mentioned that the simulation appeared realistic. 
However, differences in functional fidelity do not necessarily explain the statistically significant difference in overall task load between Realism conditions (see ~\cref{fig:overalltctsustlxplot} (Middle)). 
A detailed analysis shows (see~\cref{fig:boxplots_overall}) that mental demand and stress were significantly lower for the \condXRP{} condition, which indicates that participants perceived the task differently for the simulated device. 
Participants stated that while navigating the device using the surrogate mode, they had to rely on the real device navigating autonomously without colliding with the environment. 
As collisions do not have negative consequences for a virtual device in the \condXRP{} condition, participants potentially had a higher mental demand and stress due to the fear of potential collisions in the \condREAL{} condition.
Three participants even compensated for this perceived lack of control by waiting for the self-driving robot to catch up to the surrogate before continuing the task. 
This explains 5/7 TCT outliers in~\cref{fig:overalltctsustlxplot} (Left) and 6/8 outliers in~\cref{fig:overalltctsustlxplot} (Middle).
One participant was generally slower (2/7, 2/8 outlier).
Overall, we accept H2 as results showed differences between the Realism conditions.
 
\textbf{H3.} 
Similar to previous work~\cite{bruno2010product, faust2019mixed,Tran2023}, we found comparable qualitative feedback between both Realism conditions, as seen by the similar rankings in both conditions (see ~\cref{fig:preference}). 
However, our focus was not finding similar qualitative feedback, but determining relative validity for objective performance measures. 
We found the same order relationship for \condXRP{} and \condREAL{} in terms of \ac{TCT} between conditions \condSUR{} and \condSTE{}, as well as \condDIR{}. 
However, while SUS and TLX were following the same order relationship in both Realism conditions as well, these differences were not statistically significant. 
We partially accept H3, as, similar to H1, the \condSUR{} visualization has a beneficial effect on the user experience and follows the same order relationship in the Realism conditions.



\subsection{Insights}
We derive the following insights for \ac{XRP} validation studies, and discuss the effect of the different latency visualization methods.

\textbf{Relative Validity for Validation Studies. }
Previous work demonstrated that achieving absolute validity in \ac{XRP} validation studies is potentially out of reach~\cite{Pluemer2023, Auer2021,Mathis2021}. Our results, with respect to \ac{TCT} (H2), further emphasize this point. Therefore, with H3, we explored the notion of relative validity and demonstrated that objective performance measures can indeed have the same order relationships when comparing the real use case against an \ac{XR} simulation. In our case, the surrogate visualization outperformed the other visualization conditions when interacting with the real device and with the simulated device. Hence, as proposed by~\cite{Pluemer2023}, in order to move beyond collecting only qualitative feedback in a simulation, demonstrating relative validity  may be a path towards the validation of \ac{XRP} for collecting objective performance criteria during product development.

\textbf{Human Behavior in \ac{XRP} Framework. }
We based our experiment on a previously proposed \ac{XRP} validation framework~\cite{Pluemer2023}. 
The framework supported the design of the presented validation study and allowed us to control and discuss for potential confounding factors. 
However, while we expected differences in performance when interacting with the real device compared to the simulated device, the differences may not be explainable within the currently proposed validation framework. 
Based on the validation framework, differences could be attributed to functional fidelity being different for the simulated device (see~\cref{tab:fidelity_analysis}). 
However, subjective feedback collected from participants indicates that participant's attitude towards the real and simulated device differ, as stress and mental demand were higher when controlling the real device. 
Qualitative feedback of participants sheds some light as participants were worried of collisions between the device and the environment, particularly in the \condSUR{} condition, where the ring drove autonomously, following the pre-determined path. 
Hence, participants also ranked \condSUR{} in the \condREAL{} condition more often in third place, as 
such collisions have potentially negative consequences when happening to the real device rather than the virtual recreation. Interestingly, none of the participants who slowed down the robot during interaction and consequently produced outliers in \ac{TCT} (see~Fig.~\ref{fig:overalltctsustlxplot}) mentioned  insecurities when steering the robot. Apparently, participants who adapted their strategy and slowed down the robot may have had reduced perceived risk during the task, which may have had a positive impact on the ranking.
Also, previous work~\cite{Pluemer2023} already mentioned that in a use case, where users controlled a drone, perceived risk could have influenced objective performance. Therefore, we suggest expanding the \ac{XRP} validation framework by a discussion of factors that can potentially lead to behavioral or attitudinal differences between the real and simulated use case. Based on this discussion, appropriate data collection methods can be introduced to test for the identified factors. Identifying potential influences can also inform the characteristics of the participant sample, e.g., as the level of expertise of participants could reduce stress levels that impact performance~\cite{PREWETT2010840}
or influences user trust and attitude \cite{Sanders17}.
  
\textbf{Visualizing Processes improves Responsiveness. }
While the visualization of the robot's invisible processes, e.g., the turning wheels, did not objectively improve performance, 50\% of participants commented on the robot being more responsive with the visualization of the wheels. Furthermore, ranking data shows that the robot without any supporting latency visualization was the least preferred condition. This is in line with previous work from user interfaces that utilized progress bars~\cite{Harrison2007,Myers1985}, or countdowns~\cite{Ghafurian2020} to reduce subjective wait times. Hence, visualizing such processes using, e.g., an embedded data visualization~\cite{Willett2017}, can improve perceived responsiveness when interacting with heavy machinery.

\textbf{Sense of Agency and Trust}
Interacting with the surrogate visualization made interaction objectively faster as the motion of the surrogate was decoupled from the real device and, thus, generally faster.
However, participants stated that this also led to insecurities as they only had indirect control of the self-driving robot.
One solution is to allow users more control over the robot by modifying or confirming the robot's path~\cite{Walker2023}. This potentially improves the \ac{SoA}, which is defined as the feeling of influencing the world through controlling one's own actions \cite{Haggard12, Moore16 }, which can increase the user experience, performance and system acceptability as well \cite{Limerick14, Vantrepotte22}.
An alternative strategy can improve autonomy and decision-making capabilities of the robot, thus, delegating even more control to the robot. 
However, while delegating control can reduce cognitive load, excessive automation can negatively impact \ac{SoA}~\cite{ueda21}.
Moreover, delegating control also requires trust~\cite{Frering23} in the autonomous robot. 
In our study, a lack of trust in the autonomous system could have lead to a higher stress and mental demand when controlling the real robot due to fear of collisions and damages compared to the simulated robot.


\section{Conclusion and Future Work}
We presented an \ac{XRP} validation study that compared the performance of users steering a mobile imaging robot against them steering a virtual representation. The study compared latency visualization methods showing that they improve the user experience in human-robot interaction. Participants performed a meaningful task in a realistic setting, in our case steering a medical robot in a surgery theater. By utilizing a task that is representative for other self-driving and human-steered robots, we ensure generalizability to other forms of robots.
Our goal was to demonstrate that with detailed preparation of a validation study, we could control for confounding factors to clearly establish that achieving relative validity with \ac{XRP} methods is a feasible goal. 
For our feasibility study, we aligned all parameters of the comparative study with only functional fidelity being different between real and simulated device. However, we also emphasize that the framework potentially requires expansion beyond its current state so that it also includes a discussion of differences in expected human behavior when users interact with real and simulated devices that could influence objective and subjective data and lead to potential differences.

Performing well-structured validation studies can further expand the knowledge of the use of \ac{XR} in product development, so that more advanced stages of the product development cycle could iterate over device variations utilizing only \ac{XR} simulation hardware and approximations or smaller parts of the real device. For instance, for the presented case of prototyping supportive visualizations for a mobile imaging robot, the \ac{VST} \ac{HMD} and the control panel hardware may be sufficient for usability and performance evaluations, which would greatly decrease cost for product iterations. Evaluations could also be performed independently of the deployment location, increasing flexibility and allowing for a wider range of participants to take part in evaluations. Further validation studies are required to push the boundaries of \ac{XR} prototyping. 



\acknowledgments{%
    This work was supported by medPhoton GmbH, the TU Munich and a grant from the Austrian Research Promotion Agency (grant no. 891105).
    The evaluation was supported by the ForNero Project (BFS, AZ-1592-23) and Prof. Dr. Dirk Wilhelm, TU Munich.    
}

\bibliographystyle{bibstyle/abbrv-doi-hyperref}

\balance
\bibliography{main}

\begin{thebibliography}{10}

\bibitem{alexander2005gaming}
A.~L. Alexander, T.~Bruny{\'e}, J.~Sidman, S.~A. Weil, et~al.
\newblock {From Gaming to Training: A Review of Studies on Fidelity, Immersion, Presence, and Buy-In and Their Effects on Transfer in PC-Based Simulations and Games}.
\newblock {\em DARWARS Training Impact Group}, 5:1--14, 2005.

\bibitem{andersen2015measuring}
T.~T. Andersen, H.~B. Amor, N.~A. Andersen, and O.~Ravn.
\newblock {Measuring and Modelling Delays in Robot Manipulators for Temporally Precise Control Using Machine Learning}.
\newblock In {\em IEEE ICMLA'15}, pp. 168--175, 2015.

\bibitem{andress2018fly}
S.~Andress, A.~Johnson, M.~Unberath, A.~F. Winkler, K.~Yu, J.~Fotouhi, S.~Weidert, G.~Osgood, and N.~Navab.
\newblock {On-the-Fly Augmented Reality for Orthopedic Surgery Using a Multimodal Fiducial}.
\newblock {\em Journal of Medical Imaging}, 5(2):021209--021209, 2018.

\bibitem{anvari2005impact}
M.~Anvari, T.~Broderick, H.~Stein, T.~Chapman, M.~Ghodoussi, D.~W. Birch, C.~Mckinley, P.~Trudeau, S.~Dutta, and C.~H. Goldsmith.
\newblock {The Impact of Latency on Surgical Precision and Task Completion During Robotic-Assisted Remote Telepresence Surgery}.
\newblock {\em Computer Aided Surgery}, 10(2):93--99, 2005.

\bibitem{Auer2021}
S.~Auer, J.~Gerken, H.~Reiterer, and H.-C. Jetter.
\newblock {Comparison Between Virtual Reality and Physical Flight Simulators for Cockpit Familiarization}.
\newblock In {\em Proceedings of Mensch Und Computer 2021}, p. 378–392, 2021.

\bibitem{BARBIERI2013}
L.~Barbieri, A.~Angilica, F.~Bruno, and M.~Muzzupappa.
\newblock {Mixed Prototyping With Configurable Physical Archetype for Usability Evaluation of Product Interfaces}.
\newblock {\em Computers in Industry}, 64(3):310--323, 2013.

\bibitem{Baricevic2012}
D.~Baričević, C.~Lee, M.~Turk, T.~Höllerer, and D.~A. Bowman.
\newblock {A Hand-Held AR Magic Lens With User-Perspective Rendering}.
\newblock In {\em IEEE ISMAR}, pp. 197--206, 2012.

\bibitem{Bejczy1990}
A.~Bejczy, W.~Kim, and S.~Venema.
\newblock {The Phantom Robot: Predictive Displays for Teleoperation With Time Delay}.
\newblock In {\em Proceedings., IEEE International Conference on Robotics and Automation}, pp. 546--551 vol.1, 1990.

\bibitem{Bowman2007}
D.~A. Bowman and R.~P. McMahan.
\newblock {Virtual Reality: How Much Immersion Is Enough?}
\newblock {\em Computer}, 40(7):36--43, 2007.

\bibitem{bowman2012evaluating}
D.~A. Bowman, C.~Stinson, E.~D. Ragan, S.~Scerbo, T.~H{\"o}llerer, C.~Lee, R.~P. McMahan, and R.~Kopper.
\newblock {Evaluating Effectiveness in Virtual Environments With MR Simulation}.
\newblock In {\em Interservice/Industry Training, Simulation, and Education Conference}, vol.~4, p.~44, 2012.

\bibitem{Brooke96sus}
J.~Brooke.
\newblock {SUS: A Quick and Dirty Usability Scale}.
\newblock {\em Usability evaluation in industry}, 189(194):4--7, 1996.

\bibitem{bruno2013reliable}
F.~Bruno, A.~Angilica, F.~Cosco, and M.~Muzzupappa.
\newblock {Reliable Behaviour Simulation of Product Interface in Mixed Reality}.
\newblock {\em Engineering with Computers}, 29:375--387, 2013.

\bibitem{bruno2010functional}
F.~Bruno, A.~Angilica, F.~Cosco, M.~Muzzupappa, I.~Horvath, F.~Mandorli, and Z.~Rusak.
\newblock {Functional Behaviour Simulation of Industrial Products in Virtual Reality}.
\newblock In {\em International symposium on Tools and Methods of competitive engineering}, vol.~2, 2010.

\bibitem{bruno2010product}
F.~Bruno and M.~Muzzupappa.
\newblock {Product Interface Design: A Participatory Approach Based on Virtual Reality}.
\newblock {\em International journal of human-computer studies}, 68(5):254--269, 2010.

\bibitem{Burova2020}
A.~Burova, J.~M\"{a}kel\"{a}, J.~Hakulinen, T.~Keskinen, H.~Heinonen, S.~Siltanen, and M.~Turunen.
\newblock {Utilizing VR and Gaze Tracking to Develop AR Solutions for Industrial Maintenance}.
\newblock In {\em CHI Conf. on Human Factors in Computing Systems}, p. 1–13, 2020.

\bibitem{Chen2007}
J.~Y.~C. Chen, E.~C. Haas, and M.~J. Barnes.
\newblock {Human Performance Issues and User Interface Design for Teleoperated Robots}.
\newblock {\em IEEE Transactions on Systems, Man, and Cybernetics, Part C (Applications and Reviews)}, 37(6):1231--1245, 2007.

\bibitem{Cheng2021}
Y.~Cheng, Y.~Yan, X.~Yi, Y.~Shi, and D.~Lindlbauer.
\newblock {SemanticAdapt: Optimization-Based Adaptation of Mixed Reality Layouts Leveraging Virtual-Physical Semantic Connections}.
\newblock In {\em ACM Symp. on User Interf. Softw. and Techn.}, p. 282–297, 2021.

\bibitem{falanga2019fast}
D.~Falanga, S.~Kim, and D.~Scaramuzza.
\newblock {How Fast Is Too Fast? The Role of Perception Latency in High-Speed Sense and Avoid}.
\newblock {\em IEEE Robotics and Automation Letters}, 4(2):1884--1891, 2019.

\bibitem{farajiparvar2020brief}
P.~Farajiparvar, H.~Ying, and A.~Pandya.
\newblock {A Brief Survey of Telerobotic Time Delay Mitigation}.
\newblock {\em Frontiers in Robotics and AI}, 7:578805, 2020.

\bibitem{faust2019mixed}
F.~G. Faust, T.~Catecati, I.~de~Souza~Sierra, F.~S. Araujo, A.~R.~G. Ram{\'\i}rez, E.~M. Nickel, and M.~G. Gomes~Ferreira.
\newblock {Mixed Prototypes for the Evaluation of Usability and User Experience: Simulating an Interactive Electronic Device}.
\newblock {\em Virtual Reality}, 23:197--211, 2019.

\bibitem{Frering23}
L.~Frering, C.~Koenczoel, J.~A. Mosbacher, M.~Kames, M.~Eder, P.~Mohr-Ziak, S.~Gabl, D.~Kalkofen, D.~Albert, B.~Kubicek, and G.~Steinbauer-Wagner.
\newblock {Impact of Robot Related User Pre-experience on Cognitive Load, Trust, Trustworthiness and Satisfaction with VR Interfaces}.
\newblock In T.~Petri{\v{c}}, A.~Ude, and L.~{\v{Z}}lajpah, eds., {\em Advances in Service and Industrial Robotics}, pp. 123--131. Springer Nature Switzerland, Cham, 2023.

\bibitem{Fritz2012}
C.~O. Fritz, P.~E. Morris, and J.~J. Richler.
\newblock {Effect Size Estimates: Current Use, Calculations, and Interpretation}.
\newblock {\em Journal of Experimental Psychology: General}, 141(1):2--18, 2012.

\bibitem{fuchs2014immersive}
H.~Fuchs, A.~State, and J.-C. Bazin.
\newblock {Immersive 3D telepresence}.
\newblock {\em Computer}, 47(7):46--52, 2014.

\bibitem{gasques2021artemis}
D.~Gasques, J.~G. Johnson, T.~Sharkey, Y.~Feng, R.~Wang, Z.~R. Xu, E.~Zavala, Y.~Zhang, W.~Xie, X.~Zhang, et~al.
\newblock {Artemis: A Collaborative Mixed-Reality System for Immersive Surgical Telementoring}.
\newblock In {\em ACM CHI'21}, pp. 1--14, 2021.

\bibitem{Ghafurian2020}
M.~Ghafurian, D.~Reitter, and F.~E. Ritter.
\newblock {Countdown Timer Speed: A Trade-off between Delay Duration Perception and Recall}.
\newblock {\em ACM Trans. Comput.-Hum. Interact.}, 27(2), mar 2020.

\bibitem{Haggard12}
P.~Haggard and V.~Chambon.
\newblock Sense of agency.
\newblock {\em Current Biology}, 22(10):R390--R392, 2012.

\bibitem{Harrison2007}
C.~Harrison, B.~Amento, S.~Kuznetsov, and R.~Bell.
\newblock {Rethinking the Progress Bar}.
\newblock In {\em ACM UIST'07}, p. 115–118. Association for Computing Machinery, New York, NY, USA, 2007.

\bibitem{HART1988}
S.~G. Hart and L.~E. Staveland.
\newblock {Development of NASA-TLX (Task Load Index): Results of Empirical and Theoretical Research}.
\newblock In P.~A. Hancock and N.~Meshkati, eds., {\em Human Mental Workload}, vol.~52 of {\em Advances in Psychology}, pp. 139--183. North-Holland, 1988.

\bibitem{hashimoto2011touchme}
S.~Hashimoto, A.~Ishida, M.~Inami, and T.~Igarashi.
\newblock {TouchMe: An Augmented Reality Based Remote Robot Manipulation}.
\newblock In {\em The 21st International Conference on Artificial Reality and Telexistence, Proceedings of ICAT2011}, vol.~2, 2011.

\bibitem{Hohenstein2016}
J.~Hohenstein, H.~Khan, K.~Canfield, S.~Tung, and R.~Perez~Cano.
\newblock {Shorter Wait Times: The Effects of Various Loading Screens on Perceived Performance}.
\newblock In {\em ACM CHI EA'16}, p. 3084–3090, 2016.

\bibitem{Jensen2015}
L.~C. Jensen, K.~Fischer, D.~Shukla, and J.~Piater.
\newblock {Negotiating Instruction Strategies during Robot Action Demonstration}.
\newblock In {\em HRI'15 Extended Abstracts}, p. 143–144. ACM, 2015.

\bibitem{Jetter2020}
H.-C. Jetter, R.~R\"{a}dle, T.~Feuchtner, C.~Anthes, J.~Friedl, and C.~N. Klokmose.
\newblock {``In VR, Everything is Possible!'': Sketching and Simulating Spatially-Aware Interactive Spaces in Virtual Reality}.
\newblock CHI '20, p. 1–16, 2020.

\bibitem{Jung2018}
J.~Jung, H.~Lee, J.~Choi, A.~Nanda, U.~Gruenefeld, T.~Stratmann, and W.~Heuten.
\newblock {Ensuring Safety in Augmented Reality from Trade-off Between Immersion and Situation Awareness}.
\newblock ISMAR '18, pp. 70--79, 2018.

\bibitem{Lee2010}
C.~Lee, S.~Bonebrake, D.~A. Bowman, and T.~Höllerer.
\newblock {The Role of Latency in the Validity of AR Simulation}.
\newblock In {\em 2010 IEEE Virtual Reality Conference (VR)}, pp. 11--18, 2010.

\bibitem{Lee2009}
C.~Lee, S.~Bonebrake, T.~Hollerer, and D.~A. Bowman.
\newblock {A Replication Study Testing the Validity of AR Simulation in VR for Controlled Experiments}.
\newblock ISMAR '09, pp. 203--204, 2009.

\bibitem{Lee2013}
C.~Lee, G.~A. Rincon, G.~Meyer, T.~Höllerer, and D.~A. Bowman.
\newblock {The Effects of Visual Realism on Search Tasks in Mixed Reality Simulation}.
\newblock {\em IEEE TVCG}, 19(4):547--556, 2013.

\bibitem{Limerick14}
H.~Limerick, D.~Coyle, and J.~W. Moore.
\newblock The experience of agency in human-computer interactions: a review.
\newblock {\em Frontiers in Human Neuroscience}, 8, 2014.

\bibitem{Feiyu2022}
F.~Lu and Y.~Xu.
\newblock {Exploring Spatial UI Transition Mechanisms with Head-Worn Augmented Reality}.
\newblock CHI '22, 2022.

\bibitem{luz2023enhanced}
R.~Luz, J.~L. Silva, and R.~Ventura.
\newblock {Enhanced Teleoperation Interfaces for Multi-Second Latency Conditions: System Design and Evaluation}.
\newblock {\em IEEE Access}, 11:10935--10953, 2023.

\bibitem{MacKenzie1993}
I.~S. MacKenzie and C.~Ware.
\newblock {Lag as a Determinant of Human Performance in Interactive Systems}.
\newblock In {\em INTERACT '93 and CHI '93}, p. 488–493, 1993.

\bibitem{Makela2020}
V.~M\"{a}kel\"{a}, R.~Radiah, S.~Alsherif, M.~Khamis, C.~Xiao, L.~Borchert, A.~Schmidt, and F.~Alt.
\newblock {Virtual Field Studies: Conducting Studies on Public Displays in Virtual Reality}.
\newblock CHI '20, p. 1–15, 2020.

\bibitem{Marquardt2020}
A.~Marquardt, C.~Trepkowski, T.~D. Eibich, J.~Maiero, E.~Kruijff, and J.~Schöning.
\newblock {Comparing Non-Visual and Visual Guidance Methods for Narrow Field of View Augmented Reality Displays}.
\newblock {\em IEEE TVCG}, 26(12):3389--3401, 2020.

\bibitem{Mathis2022}
F.~Mathis, J.~O'Hagan, K.~Vaniea, and M.~Khamis.
\newblock {Stay Home! Conducting Remote Usability Evaluations of Novel Real-World Authentication Systems Using Virtual Reality}.
\newblock AVI 2022, 2022.

\bibitem{Mathis2021}
F.~Mathis, K.~Vaniea, and M.~Khamis.
\newblock {RepliCueAuth: Validating the Use of a Lab-Based Virtual Reality Setup for Evaluating Authentication Systems}.
\newblock CHI '21, 2021.

\bibitem{mcmahan2016interaction}
R.~P. McMahan, C.~Lai, and S.~K. Pal.
\newblock {Interaction Fidelity: The Uncanny Valley of Virtual Reality Interactions}.
\newblock VAMR'16, pp. 59--70. Springer, 2016.

\bibitem{Medeiros2022}
D.~Medeiros, M.~McGill, A.~Ng, R.~McDermid, N.~Pantidi, J.~Williamson, and S.~Brewster.
\newblock {From Shielding to Avoidance: Passenger Augmented Reality and the Layout of Virtual Displays for Productivity in Shared Transit}.
\newblock {\em IEEE TVCG}, 28(11):3640--3650, 2022.

\bibitem{Min2019}
X.~Min, W.~Zhang, S.~Sun, N.~Zhao, S.~Tang, and Y.~Zhuang.
\newblock {VPModel: High-Fidelity Product Simulation in a Virtual-Physical Environment}.
\newblock {\em IEEE TVCG}, 25(11):3083--3093, 2019.

\bibitem{Moore16}
J.~W. Moore.
\newblock What is the sense of agency and why does it matter?
\newblock {\em Frontiers in Psychology}, 7, 2016.

\bibitem{Muender2022}
T.~Muender, M.~Bonfert, A.~V. Reinschluessel, R.~Malaka, and T.~D\"{o}ring.
\newblock {Haptic Fidelity Framework: Defining the Factors of Realistic Haptic Feedback for Virtual Reality}.
\newblock CHI '22, 2022.

\bibitem{Myers1985}
B.~A. Myers.
\newblock {The Importance of Percent-Done Progress Indicators for Computer-Human Interfaces}.
\newblock In {\em Proceedings of the SIGCHI Conference on Human Factors in Computing Systems}, CHI '85, p. 11–17. Association for Computing Machinery, New York, NY, USA, 1985.

\bibitem{Ng2021}
A.~Ng, D.~Medeiros, M.~McGill, J.~Williamson, and S.~Brewster.
\newblock {The Passenger Experience of Mixed Reality Virtual Display Layouts in Airplane Environments}.
\newblock ISMAR'21, pp. 265--274, 2021.

\bibitem{noguera2023quantifying}
A.~Noguera~Cundar, R.~Fotouhi, Z.~Ochitwa, and H.~Obaid.
\newblock {Quantifying the Effects of Network Latency for a Teleoperated Robot}.
\newblock {\em Sensors}, 23(20):8438, 2023.

\bibitem{oberhauser2017virtual}
M.~Oberhauser and D.~Dreyer.
\newblock {A Virtual Reality Flight Simulator for Human Factors Engineering}.
\newblock {\em Cognition, Technology \& Work}, 19:263--277, 2017.

\bibitem{Pettersson2019}
I.~Pettersson, M.~Karlsson, and F.~T. Ghiurau.
\newblock {Virtually the Same Experience? Learning from User Experience Evaluation of In-Vehicle Systems in VR and in the Field}.
\newblock DIS '19, p. 463–473, 2019.

\bibitem{Pluemer2023}
J.~H. Plümer and M.~Tatzgern.
\newblock {Towards a Framework for Validating XR Prototyping for Performance Evaluations of Simulated User Experiences}.
\newblock In {\em IEEE ISMAR'23}, pp. 810--819, 2023.

\bibitem{PREWETT2010840}
M.~S. Prewett, R.~C. Johnson, K.~N. Saboe, L.~R. Elliott, and M.~D. Coovert.
\newblock {Managing Workload in Human–Robot Interaction: A Review of Empirical Studies}.
\newblock {\em Computers in Human Behavior}, 26(5):840--856, 2010.
\newblock Advancing Educational Research on Computer-supported Collaborative Learning (CSCL) through the use of gStudy CSCL Tools.

\bibitem{qian2017towards}
L.~Qian, M.~Unberath, K.~Yu, B.~Fuerst, A.~Johnson, N.~Navab, and G.~Osgood.
\newblock {Towards Virtual Monitors for Image Guided Interventions-Real-Time Streaming to Optical See-Through Head-Mounted Displays}.
\newblock {\em arXiv preprint arXiv:1710.00808}, 2017.

\bibitem{qian2019review}
L.~Qian, J.~Y. Wu, S.~P. DiMaio, N.~Navab, and P.~Kazanzides.
\newblock {A Review of Augmented Reality in Robotic-Assisted Surgery}.
\newblock {\em IEEE Transactions on Medical Robotics and Bionics}, 2(1):1--16, 2019.

\bibitem{Ragan2009}
E.~Ragan, C.~Wilkes, D.~A. Bowman, and T.~Hollerer.
\newblock {Simulation of Augmented Reality Systems in Purely Virtual Environments}.
\newblock VR'09, pp. 287--288, 2009.

\bibitem{Rakita2020}
D.~Rakita, B.~Mutlu, and M.~Gleicher.
\newblock {Effects of Onset Latency and Robot Speed Delays on Mimicry-Control Teleoperation}.
\newblock In {\em ACM/IEEE HRI '20}, p. 519–527, 2020.

\bibitem{Ren2016}
D.~Ren, T.~Goldschwendt, Y.~Chang, and T.~Höllerer.
\newblock {Evaluating Wide-Field-of-View Augmented Reality With Mixed Reality Simulation}.
\newblock VR'16, pp. 93--102, 2016.

\bibitem{sachdeva2021using}
N.~Sachdeva, M.~Klopukh, R.~S. Clair, and W.~E. Hahn.
\newblock {Using Conditional Generative Adversarial Networks To Reduce the Effects of Latency in Robotic Telesurgery}.
\newblock {\em Journal of Robotic Surgery}, 15:635--641, 2021.

\bibitem{Sanders17}
T.~L. Sanders, K.~MacArthur, W.~Volante, G.~Hancock, T.~MacGillivray, W.~Shugars, and P.~A. Hancock.
\newblock Trust and prior experience in human-robot interaction.
\newblock {\em Proceedings of the Human Factors and Ergonomics Society Annual Meeting}, 61(1):1809--1813, 2017.

\bibitem{Savino2019}
G.-L. Savino, N.~Emanuel, S.~Kowalzik, F.~Kroll, M.~C. Lange, M.~Laudan, R.~Leder, Z.~Liang, D.~Markhabayeva, M.~Schmei\ss{}er, N.~Sch\"{u}tz, C.~Stellmacher, Z.~Xu, K.~Bub, T.~Kluss, J.~Maldonado, E.~Kruijff, and J.~Sch\"{o}ning.
\newblock {Comparing Pedestrian Navigation Methods in Virtual Reality and Real Life}.
\newblock ICMI '19, p. 16–25, 2019.

\bibitem{Schott2023a}
D.~Schott, F.~Heinrich, L.~Stallmeister, J.~Moritz, B.~Hensen, and C.~Hansen.
\newblock Is this the vreal life? manipulating visual fidelity of immersive environments for medical task simulation.
\newblock In {\em 2023 IEEE International Symposium on Mixed and Augmented Reality (ISMAR)}, pp. 1171--1180. IEEE, 2023.

\bibitem{tang2018augmented}
R.~Tang, L.-F. Ma, Z.-X. Rong, M.-D. Li, J.-P. Zeng, X.-D. Wang, H.-E. Liao, and J.-H. Dong.
\newblock {Augmented Reality Technology for Preoperative Planning and Intraoperative Navigation During Hepatobiliary Surgery: A Review of Current Methods}.
\newblock {\em Hepatobiliary \& Pancreatic Diseases International}, 17(2):101--112, 2018.

\bibitem{Song2023}
U.~E. Tianyu~Song, Kevin~Yu and N.~Navab.
\newblock {Augmented Reality Collaborative Medical Displays (ARC-MeDs) for Multi-User Surgical Planning and Intra-Operative Communication}.
\newblock {\em Computer Methods in Biomechanics and Biomedical Engineering: Imaging \& Visualization}, 11(4):1042--1049, 2023.

\bibitem{Tran2023}
T.~T.~M. Tran, C.~Parker, M.~Hoggenmüller, L.~Hespanhol, and M.~Tomitsch.
\newblock {Simulating Wearable Urban Augmented Reality Experiences in VR: Lessons Learnt from Designing Two Future Urban Interfaces}.
\newblock {\em Multim. Tech. and Intera.}, 7(2), 2023.

\bibitem{ueda21}
S.~Ueda, R.~Nakashima, and T.~Kumada.
\newblock Influence of levels of automation on the sense of agency during continuous action.
\newblock {\em Scientific reports}, 11(1):2436, 2021.

\bibitem{Vantrepotte22}
Q.~Vantrepotte, B.~Berberian, M.~Pagliari, and V.~Chambon.
\newblock Leveraging human agency to improve confidence and acceptability in human-machine interactions.
\newblock {\em Cognition}, 222:105020, 2022.

\bibitem{Voit2019}
A.~Voit, S.~Mayer, V.~Schwind, and N.~Henze.
\newblock {Online, VR, AR, Lab, and In-Situ: Comparison of Research Methods to Evaluate Smart Artifacts}.
\newblock CHI '19, p. 1–12, 2019.

\bibitem{Walker2023}
M.~Walker, T.~Phung, T.~Chakraborti, T.~Williams, and D.~Szafir.
\newblock {Virtual, Augmented, and Mixed Reality for Human-robot Interaction: A Survey and Virtual Design Element Taxonomy}.
\newblock {\em J. Hum.-Robot Interact.}, 12(4), jul 2023.

\bibitem{walker2019robot}
M.~E. Walker, H.~Hedayati, and D.~Szafir.
\newblock {Robot Teleoperation With Augmented Reality Virtual Surrogates}.
\newblock In {\em ACM/IEEE HRI'19}, pp. 202--210, 2019.

\bibitem{Watson2022}
K.~Watson, R.~Bretin, M.~Khamis, and F.~Mathis.
\newblock {The Feet in Human-Centred Security: Investigating Foot-Based User Authentication for Public Displays}.
\newblock CHI EA '22, 2022.

\bibitem{Weiss2021}
M.~Weiß, K.~Angerbauer, A.~Voit, M.~Schwarzl, M.~Sedlmair, and S.~Mayer.
\newblock {Revisited: Comparison of Empirical Methods to Evaluate Visualizations Supporting Crafting and Assembly Purposes}.
\newblock {\em IEEE TVCG}, 27(2):1204--1213, 2021.

\bibitem{Willett2017}
W.~Willett, Y.~Jansen, and P.~Dragicevic.
\newblock {Embedded Data Representations}.
\newblock {\em IEEE TVCG}, 23(1):461--470, 2017.

\bibitem{Wobbrock2011}
J.~O. Wobbrock, L.~Findlater, D.~Gergle, and J.~J. Higgins.
\newblock {The Aligned Rank Transform for Nonparametric Factorial Analyses Using Only Anova Procedures}.
\newblock CHI'11, p. 143–146, 2011.

\bibitem{xu2014determination}
S.~Xu, M.~Perez, K.~Yang, C.~Perrenot, J.~Felblinger, and J.~Hubert.
\newblock {Determination of the Latency Effects on Surgical Performance and the Acceptable Latency Levels in Telesurgery Using the dV-Trainer{\textregistered} Simulator}.
\newblock {\em Surgical endoscopy}, 28:2569--2576, 2014.

\bibitem{Yu2023}
K.~Yu, D.~Roth, R.~Strak, F.~Pankratz, J.~Reichling, C.~Kraetsch, S.~Weidert, M.~Lazarovici, N.~Navab, and U.~Eck.
\newblock {Mixed Reality 3D Teleconsultation for Emergency Decompressive Craniotomy: An Evaluation with Medical Residents}.
\newblock In {\em IEEE ISMAR'23}, pp. 662--671, 2023.

\end{thebibliography}


\end{document}